\documentclass[12pt]{iopart}

\usepackage{iopams}  

\expandafter\let\csname equation*\endcsname\relax
\expandafter\let\csname endequation*\endcsname\relax

\usepackage{epsfig, amsmath, amssymb}
\newtheorem{theorem}{Theorem}
\newtheorem{prop}{Proposition}
\newtheorem{example}{Example}	%
\newtheorem{remark}{Remark}		%
\newtheorem{lemma}{Lemma}
\newtheorem{corollary}[example]{Corollary}

\newtheorem{definition}{Definition}	%
\newtheorem{assumption}{Assumption}
\usepackage{hyperref}
\usepackage{lineno}

\usepackage{xcolor}
\usepackage{booktabs}
\usepackage{multirow}
\usepackage{cite}

\begin{document}

\title[Random-Matrix-Induced Simplicity Bias in Over-parameterized VQCs]{Random-Matrix-Induced Simplicity Bias in Over-parameterized Variational Quantum Circuits}

\author{Jun Qi$^{1*}$, Chao-Han Huck Yang$^{2}$, Pin-Yu Chen$^{3}$, Min-Hsiu Hsieh$^{4*}$}

\address{1. School of Electrical and Computer Engineering, Georgia Institute of Technology, Atlanta, GA 30332, USA     \\ 
2. NVIDIA Research, Santa Clara, CA 95051, USA	 \\
3. IBM Thomas J. Watson Research Center, NY, 10598, USA    \\
4. Hon Hai (Foxconn) Quantum Computing Research Center, Taipei, 114, Taiwan   \\
}
\ead{jqi41@gatech.edu, min-hsiu.hsieh@foxconn.com}
\hspace{25mm}\small{* denotes corresponding authors}
\vspace{10pt}


\begin{abstract}
Over-parameterization is commonly used to increase the expressivity of variational quantum circuits (VQCs), yet deeper and more highly parameterized circuits often exhibit poor trainability and limited generalization. In this work, we provide a theoretical explanation for this phenomenon from a function-class perspective. We show that sufficiently expressive, unstructured variational ansätze enter a Haar-like universality class in which both observable expectation values and parameter gradients concentrate exponentially with system size. As a consequence, the hypothesis class induced by such circuits collapses with high probability to a narrow family of near-constant functions, a phenomenon we term simplicity bias, with barren plateaus arising as a consequence rather than the root cause. Using tools from random matrix theory and concentration of measure, we rigorously characterize this universality class and establish uniform hypothesis-class collapse over finite datasets. We further show that this collapse is not unavoidable: tensor-structured VQCs, including tensor-network-based and tensor-hypernetwork parameterizations, lie outside the Haar-like universality class. By restricting the accessible unitary ensemble through bounded tensor rank or bond dimension, these architectures prevent concentration of measure, preserve output variability for local observables, and retain non-degenerate gradient signals even in over-parameterized regimes. Together, our results unify barren plateaus, expressivity limits, and generalization collapse under a single structural mechanism rooted in random-matrix universality, highlighting the central role of architectural inductive bias in variational quantum algorithms. 
\end{abstract}

%
%
%
%
%

\section{Introduction}
\label{sec1}

Variational quantum circuits (VQCs) constitute the core computational model underlying a wide range of quantum algorithms~\cite{preskill2018quantum, bharti2022noisy}, including variational quantum eigensolvers~\cite{kandala2017hardware, zhang2022variational}, quantum approximate optimization algorithms~\cite{zhou2020quantum, wang2018quantum, zhu2022adaptive}, and quantum machine learning models~\cite{cerezo2021variational, qi2025tensorhyper, dunjko2022quantum, chen2021end, chen2022quantumCNN}. In these settings, increasing circuit depth and parameter count is often viewed as a natural route to improving expressivity and learning performance~\cite{liu2024towards, biamonte2017quantum, schuld2015introduction, huggins2019towards}. However, extensive empirical evidence has shown that deeper and more highly parameterized VQCs often suffer from severe training issues, including vanishing gradients, flat loss landscapes, and poor generalization~\cite{steffen2011quantum, resch2021benchmarking, du2021quantum}, even in the absence of optimization noise and hardware imperfections~\cite{cerezo2022challenges, kandala2017hardware, steffen2011quantum, shukla2020complete}.

A prominent line of work attributes these failures to barren plateaus~\cite{larocca2025barren, martin2023barren, holmes2022connecting, cerezo2021cost}, in which the gradients of typical cost functions vanish exponentially with system size. While this gradient-centric viewpoint has been highly influential, it leaves open a more fundamental question: what class of functions do over-parameterized variational circuits actually represent? In particular, it remains unclear whether the observed optimization difficulties arise solely from unfavorable loss landscapes or whether they reflect a deeper representational limitation intrinsic to the circuit ensemble itself.

In this work, we address this question by adopting a functional-class perspective on over-parameterized VQCs. Rather than focusing on a specific cost function~\cite{cerezo2021cost} or training procedure~\cite{qi2023theoretical}, we study the hypothesis class induced by a variational circuit architecture under random parameterization. We show that, in a broad, practically relevant regime, increasing expressivity via unstructured over-parameterization can lead to a collapse of functional diversity. With high probability over parameters, the circuit implements a near-constant function, largely independent of the input. We refer to this phenomenon as simplicity bias, emphasizing that it arises at the level of representation rather than at the level of optimization. 

Our analysis is grounded in the observation that sufficiently expressive, unstructured variational ansätze exhibit Haar-like typicality~\cite{paul2014random, nguyen2024theory}: their induced unitary ensembles reproduce low-order moments of the Haar measure on the unitary group. In this regime, tools from random matrix theory~\cite{tao2023topics, edelman2005random} and concentration of measure~\cite{hall1967measures, ledoux2001concentration} become applicable. We show that Haar-like typicality simultaneously leads to the concentration of both observable expectation values and parameter gradients, both of which decay exponentially with the number of qubits. As a consequence, the hypothesis class induced by such circuits collapses to a narrow family of near-constant functions, even before any learning dynamics are considered. 

\begin{figure}[t]
\centerline{\epsfig{figure=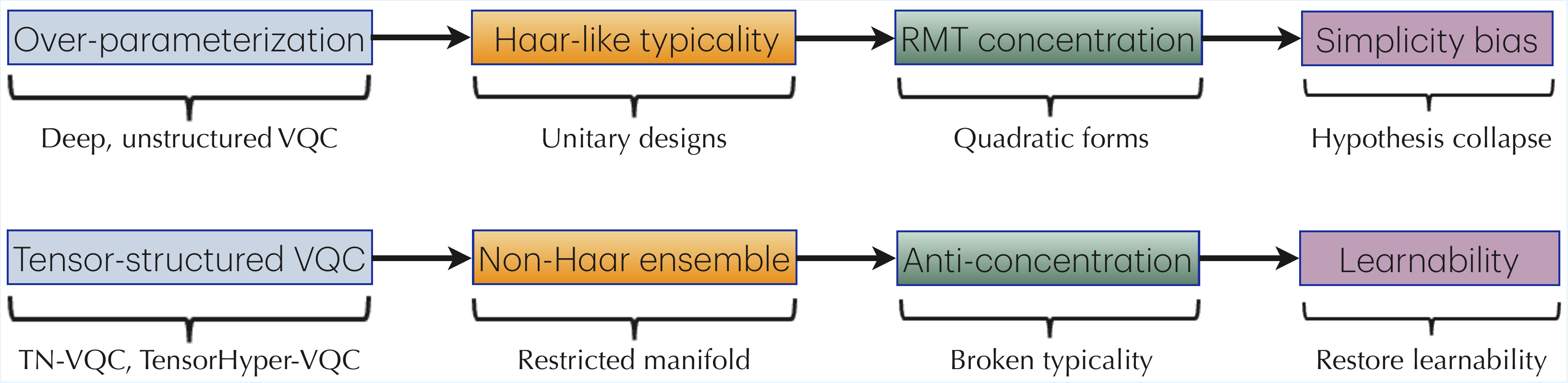, width=155mm}}
\caption{{\it Schematic illustration of simplicity bias and its mitigation in VQCs}. \textit{Top}: In sufficiently expressive, unstructured variational quantum circuits, the induced unitary ensemble approaches a random-matrix universality class, leading to concentration of observable expectation values and gradients. As a result, the hypothesis class collapses to near-constant functions, giving rise to simplicity bias and barren plateaus. \textit{Bottom}: Tensor-structured variational quantum circuits restrict the accessible unitary manifold through bounded tensor rank or bond dimension, preventing concentration of measure. This structural constraint preserves output variability and informative gradients, thereby mitigating simplicity bias even in over-parameterized regimes.}
\label{fig:framework}
\end{figure}

More importantly, this representational collapse is not universal. We demonstrate that tensor-structured variational circuits, including tensor-network-based architectures~\cite{qi2023theoretical} and tensor-hypernetwork parameterizations~\cite{qi2025tensorhyper}, lie outside the Haar-like universality class. By imposing structural constraints such as bounded tensor rank or bond dimension, these architectures restrict the accessible unitary manifold, preventing convergence to approximate unitary designs. As a result, they exhibit anti-concentration of observable expectation values and retain non-degenerate gradient signals, thereby preserving functional diversity even in highly parameterized regimes. 

As shown in Fig.~\ref{fig:framework}, the contributions of this work are threefold: 
\begin{enumerate}
\item We provide a rigorous, random-matrix-theoretic characterization of a Haar-like universality class governing over-parameterized, unstructured VQCs.
\item We show that within this regime, over-parameterization induces a representation-level collapse of the hypothesis class, referred to as simplicity bias, with barren plateaus emerging as a consequence rather than the underlying cause. 
\item We establish that tensor-structured VQCs provably escape this universality class, yielding a principled mechanism for restoring non-trivial hypothesis classes and informative learning signals. 
\end{enumerate}

Taken together, these results offer a unified structural explanation for several well-known failure modes of variational quantum algorithms and suggest that learnability is governed not only by expressivity, but by the geometry of the induced unitary ensemble. This perspective highlights the central role of architectural inductive bias in the design of scalable and trainable variational quantum algorithms.

\section{Results}
\label{sec2}

\subsection{Problem Setup and Assumptions}

\begin{figure}[t]
\centerline{\epsfig{figure=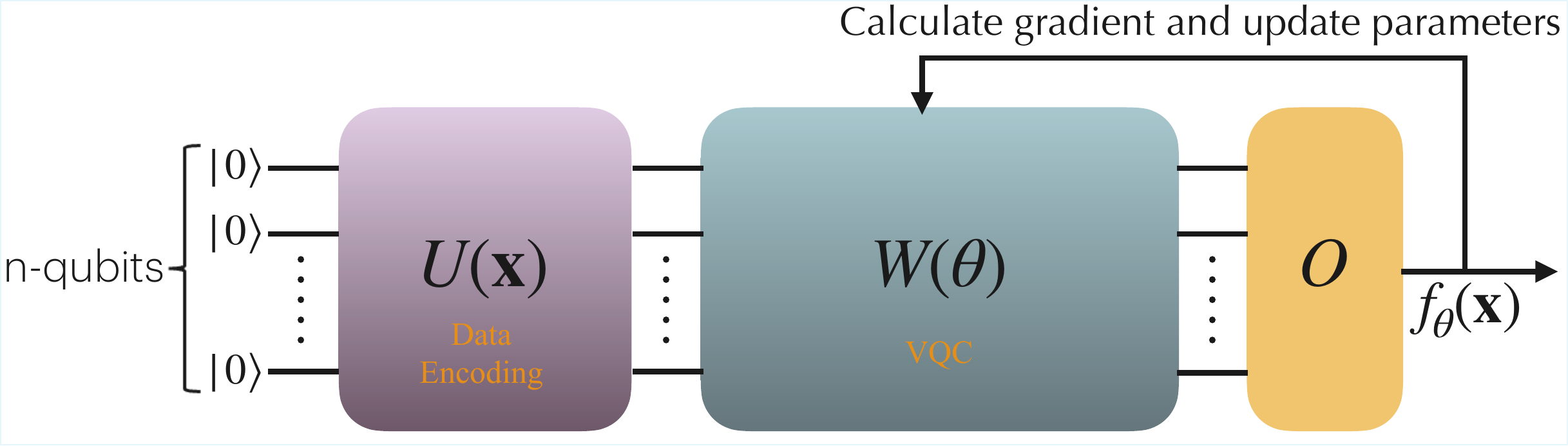, width=110mm}}
\caption{{\it Schematic of a VQC considered in this work}. A VQC consists of an input quantum state initialized as $\vert 0 \rangle^{\otimes n}$, followed by a data-encoding unitary $U(\textbf{x})$, a parameterized variational ansätze $W(\boldsymbol{\theta})$, and a measurement of a Hermitian observable $O$, yielding the scalar output $f_{\boldsymbol{\theta}}(\textbf{x})$. The expressive and statistical properties of the induced function family $\{f_{\boldsymbol{\theta}}\}$ depend on the structure of the variational ansätze $W(\boldsymbol{\theta})$.}
\label{fig:vqc}
\end{figure}

Fig.~\ref{fig:vqc} illustrates the generic VQC architecture analyzed throughout this work. The circuit maps a classical input $\textbf{x}$ to scalar output $f_{\boldsymbol{\theta}}(\textbf{x})$ by encoding data via a fixed unitary $U(\textbf{x})$, applying a parameterized unitary ansätze $W(\boldsymbol{\theta})$, and measuring a bounded observable $O$ on the resulting quantum state. Concretely, we consider an $n$-qubit VQC of the standard form: 
\begin{equation}
\label{eq:2_1}
f_{\boldsymbol{\theta}}(\textbf{x}) = \left\langle 0^{\otimes n} \left\vert U^{\dagger}(\textbf{x}) W^{\dagger}(\boldsymbol{\theta}) O W(\boldsymbol{\theta}) U(\textbf{x}) \right\vert 0^{\otimes n} \right\rangle,  
\end{equation}
where $\boldsymbol{\theta} \in \mathbb{R}^p$ denotes the circuit parameters and $O$ is a Hermitian observable satisfying $\lVert O \rVert_{2} \le 1$. Our goal is to characterize the expressivity and learnability properties of the hypothesis class induced by $f_{\boldsymbol{\theta}}$, particularly in regimes where circuit depth and parameterization scale with system size.

\begin{definition}[Haar-random]. We begin by recalling the standard notion of Haar-random. 
\label{def1}
The unitary group $U(2^n)$ admits a unique probability measure that is invariant under left and right multiplication by arbitrary unitaries, known as Haar measure. A unitary drawn from this measure is said to be Haar-random and represents maximal statistical symmetry in Hilbert space.
\end{definition}

Expectation values of bounded observables evaluated on Haar-random quantum states exhibit intense concentration of measure: as the Hilbert space dimension grows, such quantities concentrate sharply around their mean values. Throughout this work, Haar-random refers strictly to sampling from the exact Haar measure on $U(2^n)$.

In contrast, many physically motivated circuit ensembles are not exactly Haar-random but reproduce Haar moments up to finite order. These ensembles are commonly referred to as approximate unitary designs. We formalize the notion of over-parameterization via Assumption~\ref{ass1}, where we use the term Haar-like typicality to describe this approximate behavior.

\begin{assumption}[Haar-like typicality].
\label{ass1}
The variational ansätze $W(\boldsymbol{\theta})$ are sufficiently deep and expressive such that, for almost all parameter values $\boldsymbol{\theta}$, the induced distribution of unitaries $W(\boldsymbol{\theta})$ forms an approximate unitary $2$-design on $U(2^{n})$. 
\end{assumption}

In Assumption~\ref{ass1}, “almost all parameter” refers to typical draws of parameters from smooth distributions (e.g., uniform or Gaussian~\cite{zhang2022gaussian}), and over-parameterization denotes expressivity sufficient to approximate low-order Haar moments, rather than merely a large parameter count. 

This assumption captures a regime known to arise in a variety of unstructured circuit families, including hardware-efficient ansätze with all-to-all or nearest-neighbor connectivity, when circuit depth scales at least linearly with the number of qubits and parameters are initialized independently from smooth distributions. 

\vspace{1.5mm}
\textit{Scope and Interpretation of the Assumption}. Assumption~\ref{ass1} does not require exact Haar-random, nor does it assert that all circuit architectures reach this regime at finite depth. Instead, it characterizes an empirically and theoretically relevant universality class in which sufficiently expressive, unstructured variational circuits exhibit Haar-like typical behavior in low-order moments. Moreover, our results rely only on approximate design behavior sufficient to induce concentration of measure for bounded observables. The conclusions, therefore, apply whenever the circuit ensemble reproduces Haar statistics up to second order, regardless of the specific microscopic architecture. 

Besides, while Haar-like typicality underlies many known results on barren plateaus, our focus here is representational rather than algorithmic. Assumption~\ref{ass1} formalizes whether the hypothesis class induced by $f_{\boldsymbol{\theta}}$ can represent meaningfully distinct functions and whether infinitesimal parameter perturbations induce non-vanishing functional responses. The theoretical results that follow characterize the representational consequences of this typicality and are independent of any specific optimization procedure or noise model.

\subsection{Theoretical Results}

We now present the main theoretical results of this work. Theorems \ref{thm:thm1} and \ref{thm:thm2} characterize the Haar-like universality class associated with sufficiently expressive, unstructured VQCs. Propositions \ref{prop:prop2} and \ref{prop:prop3} establish that tensor-structured VQCs lie outside this universality class, and Theorem~\ref{thm:thm3} formalizes how such structure mitigates the resulting collapse.

\vspace{1.5mm}
\textit{Random-Matrix-Induced Output Concentration}. We first characterize the regime in which a variational ansatz becomes sufficiently expressive to approximate a unitary 2-design, a behavior observed in deep random circuits and hardware-efficient ansätze. Under Assumption~\ref{ass1}, for a fixed input $\textbf{x}$ and typical parameter values $\boldsymbol{\theta}$, the quantum state
\begin{equation}
\vert \boldsymbol{\psi}_{\boldsymbol{\theta}, \textbf{x}} \rangle:= W(\boldsymbol{\theta}) U(\textbf{x}) \vert 0^{\otimes n} \rangle 
\end{equation}
behaves as a Haar-random state in $\mathbb{C}^{2^{n}}$. The VQC output $f_{\boldsymbol{\theta}}(\textbf{x})$ is a quadratic form of a random vector. 

\begin{theorem}[Output concentration].
\label{thm:thm1}
Under Assumption~\ref{ass1}, for any fixed input $\textbf{x}$, 
\begin{equation}
\label{eq:3}
\mathbb{E}\left[f_{\boldsymbol{\theta}}(\textbf{x}) \right] = \frac{1}{2^{n}} \text{\rm Tr}(O), 
\end{equation}
and 
\begin{equation}
\label{eq:4}
\operatorname{Var}(f_{\boldsymbol{\theta}}(\textbf{x})) = \mathcal{O}(2^{-n}). 
\end{equation}
Moreover, for any $\epsilon > 0$, 
\begin{equation}
\label{eq:5}
\operatorname{Pr}\left( \left\vert f_{\boldsymbol{\theta}}(\textbf{x}) - \frac{1}{2^{n}}\rm{Tr}(O)\right\vert > \epsilon \right) \le 2\exp\left(-c \epsilon^{2} 2^{n}\right), 
\end{equation}
for some universal constant $c > 0$. 
\end{theorem}

Theorem~\ref{thm:thm1} shows that, in the Haar-like typical regime, deviations of the circuit output from its mean are exponentially suppressed in the Hilbert-space dimension. As a result, for almost all parameter settings, an over-parameterized VQC implements a function that is nearly constant for any fixed input.

\vspace{1mm}
\textit{Gradient Concentration and Typical Flatness}. We next show that the same random-matrix mechanism governs the behavior of parameter gradients. For typical parameterizations, the derivative of the VQC output with respect to a parameter $\theta_{k}$ can be written as:
\begin{equation}
\frac{\partial f_{\boldsymbol{\theta}}(\textbf{x})}{\partial \theta_{k}} = \langle \boldsymbol{\psi}_{\boldsymbol{\theta}, \textbf{x}} \vert G_{k} \vert \boldsymbol{\psi}_{\boldsymbol{\theta}, \textbf{x}} \rangle, 
\end{equation}
where $G_{k}$ is a bounded Hermitian operator determined by the circuit structure.

\begin{theorem}[Gradient concentration].
\label{thm:thm2}
Under Assumption~\ref{ass1}, for any fixed input $\textbf{x}$ and any circuit parameter $\theta_{k}$, 
\begin{equation}
\mathbb{E}\left[ \frac{\partial f_{\boldsymbol{\theta}}(\textbf{x})}{\partial \theta_{k}} \right] = 0, \hspace{3mm} \operatorname{Var} \left( \frac{\partial f_{\boldsymbol{\theta}}(\textbf{x})}{\partial \theta_k} \right) = \mathcal{O}(2^{-n}). 
\end{equation}
where the expectation and variance are taken over random initialization of $\boldsymbol{\theta}$. 
\end{theorem}

Theorem~\ref{thm:thm2} implies that, in the Haar-like typicality regime, gradients concentrate sharply around zero, with variance decaying exponentially in the number of qubits. Almost all parameter directions, therefore, become locally uninformative at initialization, yielding effectively flat loss landscapes. Importantly, this phenomenon arises from the representational typicality of the circuit ensemble, rather than from any specific optimization strategy, cost function, or noise mechanism.

\vspace{1mm}
\textit{Simplicity Bias and Hypothesis Class Collapse}. Theorem~\ref{thm:thm1} and \ref{thm:thm2} together imply a collapse of the effective hypothesis class induced by over-parameterized, unstructured VQCs. 

\begin{definition}[Simplicity bias]. 
\label{def2}
A hypothesis class $\mathcal{F} = \{f_{\boldsymbol{\theta}}\}$ exhibits simplicity bias if, with high probability over parameter choice, functions in $\mathcal{F}$ concentrate around a low-complexity subset of near-constant functions, regardless of parameter count. 
\end{definition}

\begin{corollary}[Uniform hypothesis-class collapse over finite datasets].
\label{cor1}
Let $\mathcal{D} = \{\textbf{x}_1, ..., \textbf{x}_{m}\}$ be a finite dataset of size $m=\operatorname{poly}(n)$. Under Assumption~\ref{ass1}, with a probability at least $1 - \delta$ over the circuit parameters,
\begin{equation}
\max\limits_{\textbf{x}_i, \textbf{x}_j \in \mathcal{D}} \left\vert f_{\boldsymbol{\theta}}(\textbf{x}_i) - f_{\boldsymbol{\theta}}(\textbf{x}_j)	\right\vert \le \epsilon, 
\end{equation}
where $\epsilon = \mathcal{O}(2^{-n/2})$ and $\delta \le m^2 \exp(-c 2^n)$. 
\end{corollary}

This corollary follows by applying Theorem~{thm:thm1} to each input and taking a union bound over all input pairs. It shows that over-parameterization alone does not guarantee expressive hypothesis classes in VQCs: in the Haar-like regime, the hypothesis class collapses uniformly to near-constant functions. 

\vspace{1mm}

\textit{Tensor-Structured VQCs and the Breakdown of Haar-like Typicality}. The simplicity bias established above relies critically on Haar-like typicality. We now show that tensor-structured VQCs lie outside this random-matrix universality class. 

Tensor-structured VQCs impose explicit structural constraints through bounded tensor rank or bond dimension, restricting the accessible unitaries to a low-dimensional subset of $U(2^n)$. We consider two representative architectures: TN-VQC and TensorHyper-VQC (Fig. \ref{fig:framework}), both of which enforce bounded entanglement and polynomial parameterization.

\begin{figure}[t]
\centerline{\epsfig{figure=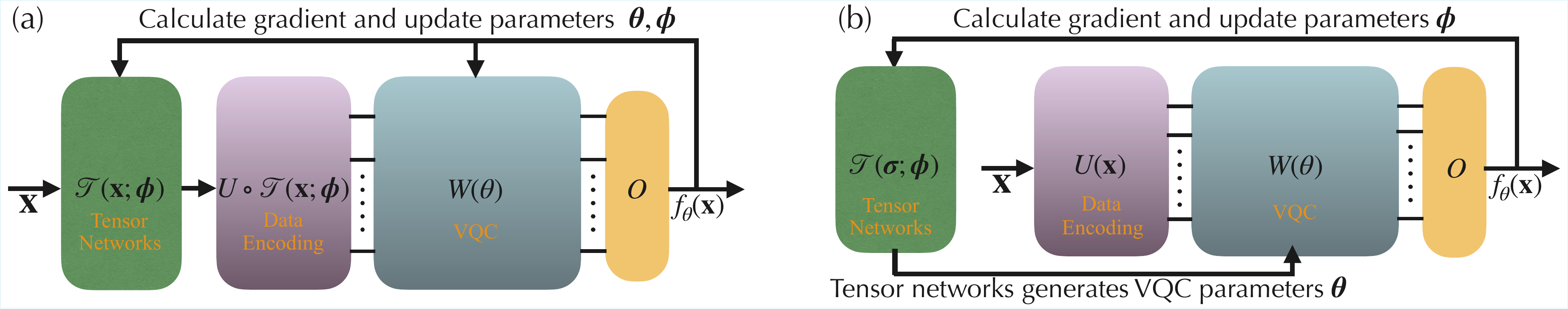, width=155mm}}
\caption{{\it Tensor-structured VQC architectures}. (a) TN-VQC: A tensor network module transforms the classical input $\textbf{x}$ into a lower-dimensional feature $\mathcal{T}(\textbf{x}; \boldsymbol{\phi})$, which is further converted into quantum state $U\circ \mathcal{T}(\textbf{x}; \boldsymbol{\phi})$ via the encoding unitary $\mathcal{T}(\textbf{x}; \boldsymbol{\phi})$. The encoded quantum state is then processed by a VQC $W(\boldsymbol{\theta})$ and measured via an observable $O$ to produce the output $f_{\boldsymbol{\theta}}(\textbf{x})$. Both the encoding parameters $\boldsymbol{\phi}$ and the circuit parameters $\boldsymbol{\theta}$ are updated through gradient-based optimization. (b) TensorHyper-VQC: A tensor-network acts as a hypernetwork that generates the variational circuit parameters $\boldsymbol{\theta}$ directly through $\mathcal{T}(\boldsymbol{\sigma}; \boldsymbol{\phi})$ by using a Gaussian random vector $\boldsymbol{\sigma}$. The data encoding is fixed, while the tensor network induces structured correlations among the parameters of $W(\boldsymbol{\theta})$. In both architectures, the tensor-network structure constrains the accessible unitary ensemble, breaking Haar-like typicality and mitigating random-matrix-induced simplicity bias.}
\label{fig:framework}
\end{figure}

\begin{prop}[Failure of Approximate Unitary Designs].
\label{prop:prop2}
Let $\mathcal{U}_{\rm ts} \in U(2^n)$ denote the ensemble of unitaries generated by a tensor-structured VQC with a fixed rank (or bond dimension) independent of $n$. Then, $\mathcal{U}_{\rm ts}$ does not form an approximate unitary $t$-design for any fixed $t\ge 2$ as $n\rightarrow \infty$. 
\end{prop}

The bounded operator Schmidt rank implied by the tensor-network structure prevents convergence to Haar moments, which require near-maximal entanglement and isotropy across bipartitions. 

\vspace{1mm}
\textit{Anti-Concentration from Bounded Entanglement}. This breakdown of Haar-like typicality can be made explicit through entanglement and variance bounds.

\begin{lemma}
\label{lem1}
Consider a tensor-structured VQC $W(\boldsymbol{\theta})$ with tensor rank $r = \mathcal{O}(1)$. For any local or few-body observable $\mathcal{O}$, the reduced density matrix on the support of $\mathcal{O}$ depends on at most $\mathcal{O}(r^2)$ effective degrees of freedom. Consequently, the variance of $\langle O \rangle$ under random parameter initialization is bounded below by a constant independent of the total number of qubits $n$.
\end{lemma}

\begin{prop}[Anti-concentration]. 
\label{prop:prop3}
Let $O$ be a bounded local or few-body observable. For a tensor-structured VQC with fixed tensor rank $r$, there exists a constant $c(r, O) > 0$, independent of the number of qubits $n$, such that 
\begin{equation}
\operatorname{Var}\left( \langle \boldsymbol{\psi}_{\boldsymbol{\theta}, \textbf{x}} \vert O \vert \boldsymbol{\psi}_{\boldsymbol{\theta}, \textbf{x}}  \rangle \right) \ge c(r, O). 
\end{equation}
\end{prop}

Unlike Haar-like VQCs, where the variance decays exponentially as $\mathcal{O}(2^{-n})$, tensor-structured VQCs retain non-vanishing variance. Observable expectation values remain sensitive to both inputs and parameters, precluding representational collapse.

\vspace{1.5mm}
\textit{Non-trivial Hypothesis Class and Learnability Proxy}. We now formalize the consequences of anti-concentration. We define the non-trivial hypothesis class in Definition~\ref{def3} and show the tensor-network VQCs' remedy in Theorem~\ref{thm:thm3} and Corollary~\ref{corr2}.

\begin{definition}[Non-trivial hypothesis class]. 
\label{def3}
Let $\mathbb{F} = \{f_{\boldsymbol{\theta}}: \mathcal{X} \rightarrow \mathbb{R}\}$ be the hypothesis class induced by a VQC. $\mathbb{F}$ is non-trivial if there exist inputs $\textbf{x},\textbf{x}' \in \mathcal{X}$ and constants $\triangle, \eta > 0$, independent of $n$, such that 
\begin{equation}
\operatorname{Pr}\left( \vert f_{\boldsymbol{\theta}}(\textbf{x}) - f_{\boldsymbol{\theta}}(\textbf{x}') \vert \ge \triangle \right) \ge \eta. 
\end{equation}

It admits a non-degenerate learning signal if there exists a parameter index $k$, an input $\textbf{x}$, and $\gamma > 0$, independent of $n$, such that 
\begin{equation}
 \operatorname{Var}\left( \frac{\partial f_{\boldsymbol{\theta}}(\textbf{x})}{\partial \theta_{k}} \right) \ge \gamma.
\end{equation}
\end{definition}

\begin{theorem}[Tensor-structured VQC mitigates simplicity bias].
\label{thm:thm3}
Let $W(\boldsymbol{\theta})$ be a tensor-structured VQC family with tensor rank bounded independent of qubit count $n$. Assume $O$ is a bounded local observable and that the data encoding maps inputs to non-identical reduced states on the support of $O$. Then: 
\begin{enumerate}
\item (Anti-concentration of outputs) There exists a constant $c_0 > 0$, independent of $n$, such that
\begin{equation}
\operatorname{Var}(f_{\boldsymbol{\theta}}(\textbf{x})) \ge c_0
\end{equation}
for some $\textbf{x} \in \mathcal{X}$.

\item (Non-degenerate gradient signal) There exists a parameter index $k$ and $c_1 > 0$, independent of $n$, such that 
\begin{equation}
\operatorname{Var}\left( \frac{\partial f_{\boldsymbol{\theta}}(\textbf{x})}{\partial \theta_k} \right) \ge c_1. 
\end{equation}
\end{enumerate}
\end{theorem}

Theorem~\ref{thm:thm3} shows that tensor-structured VQCs fundamentally alter the typical behavior of over-parameterized circuits. By breaking Haar-like typicality, the tensor-network structure prevents both output and gradient concentration, thereby ruling out the representational collapse mechanism.

We further emphasize that the nontrivial content of Theorem~\ref{thm:thm3} is not merely that tensor-structured circuits fail to approximate Haar randomness, but that this structural restriction provably enforces persistent output variability and non-degenerate gradient signals for physically relevant local observables, thereby ruling out hypothesis-class collapse even in the over-parameterized regime. 

\begin{corollary}[Restored learnability in over-parameterized settings]. 
\label{corr2}
Under the assumptions of Theorem~\ref{thm:thm3}, tensor-structured VQCs admit a non-trivial hypothesis class and a non-degenerate learning signal that persists as $n\rightarrow \infty$. 
\end{corollary}

\begin{remark}[On learnability].
\label{remark1}
Throughout this work, restored learnability refers to the absence of representational and gradient collapse due to Haar-like typicality. Non-vanishing output and gradient variance ensure functional diversity and informative descent directions at initialization, but, by themselves, do not guarantee successful training or generalization.
\end{remark}

\vspace{1.5mm}
\textit{Unified Interpretation}. The above results admit a unified interpretation: barren plateaus, expressivity collapse, and generalization failure are distinct manifestations of the same random-matrix universality class that governs over-parameterized, unstructured VQCs. Tensor-structured architectures escape this universality class by restricting the accessible unitary ensemble, thereby preventing concentration of measure and restoring functional and gradient variability.

\subsection{Numerical Simulation}

To complement our theoretical analysis, we present a minimal numerical simulation that illustrates the finite-size behavior predicted by our results. The purpose of this experiment is not empirical benchmarking, but rather to probe how variance collapse and its tensor-structured mitigation manifest at finite system sizes and finite database sizes. 

We consider three classes of VQCs: a naive unstructured VQC, a TN-VQC, and a TensorHyper-VQC, all illustrated with $n=12$ qubits and depth $L=6$. For each model, we evaluate the empirical variance $\operatorname{Var}_{\textbf{x} \sim \mathcal{D}_{m}}[f_{\boldsymbol{\theta}}(\textbf{x})]$ of the circuit output over a dataset of size $m$, where $\mathcal{D}_m$ denotes a finite sample drawn from a fixed input distribution. The dataset size $m$ is varied from $2^3$ to $2^8$, and results are averaged over multiple random initializations. 

To isolate representational effects from task-specific structure, we construct a synthetic input dataset as follows. Each input sample $\textbf{x} \in \mathbb{R}^p$ is drawn independently from a fixed isotropic distribution (standard normal), and normalized to unit norm. No labels are used, as the objective is solely to evaluate the variability of circuit outputs induced by different inputs. The dataset size varied from $2^3$ to $2^8$, and results are averaged over multiple random circuit initializations. Error bars denote one standard deviation across $10$ seeds.

\begin{figure}[t]
\centerline{\epsfig{figure=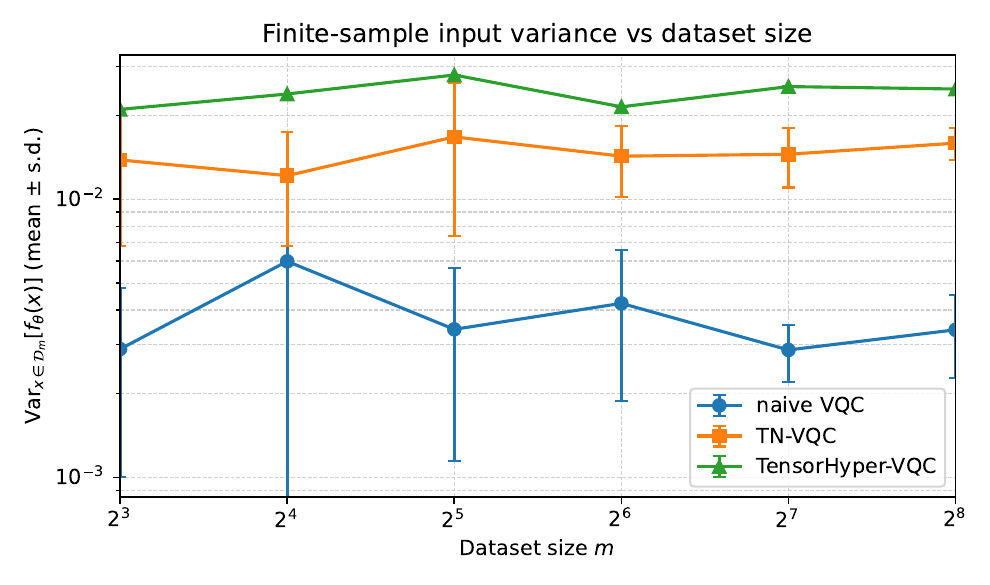, width=95mm}}
\caption{{\it Finite-sample input variance and breakdown of typicality}. Empirical variance of the model output $\operatorname{Var}_{\textbf{x} \sim \mathcal{D}_{m}}[f_{\boldsymbol{\theta}}(\textbf{x})]$ evaluated on datasets of size $m=2^3, ..., 2^8$ for a naive VQC, a TN-VQC, and a TensorHyper-VQC. Values are averaged over random initializations, with error bars denoting one standard deviation. Tensor-structured architectures exhibit stable, non-vanishing across dataset sizes, in contrast to the strongly concentrated behavior of the unstructured VQC.}
\label{fig:var}
\end{figure}

Consistent with Theorems~\ref {thm:thm1} and \ref{thm:thm3}, as shown in Figure~\ref{fig:var}, the naive VQC exhibits a pronounced concentration of output values: the empirical variance remains small and insensitive primarily to increasing dataset size, reflecting the onset of Haar-like typicality. In contrast, both tensor-structured architectures retain substantially larger and stable output variance across all dataset sizes. This behavior demonstrates a breakdown of Haar-like typicality and the persistence of nontrivial functional variability, as predicted by Proposition~\ref{prop:prop3} and Theorem~\ref{thm:thm3}. 

Notably, the qualitative separation between unstructured and tensor-structured circuits remains stable across dataset sizes, indicating that the observed effect is not an artifact of finite-sample fluctuations. Instead, it provides a finite-size illustration of the random-matrix universality principle underlying our theoretical framework. 

\section{Discussion}

This work provides a theoretical explanation for a persistent empirical phenomenon in variational quantum algorithms: increasing circuit expressivity through unstructured over-parameterization does not necessarily improve learning performance and may instead induce a collapse of functional diversity. By modeling sufficiently expressive, hardware-efficient VQCs within a Haar-like universality class, we show that both outputs and parameter gradients concentrate sharply due to concentration-of-measure effects. As a consequence, almost all parameter settings yield near-constant functions, leading to a strong simplicity bias. Importantly, this collapse is not merely an optimization pathology, but a structural property of the hypothesis class induced by the circuit ensemble. 

While our analysis is asymptotic, extensive numerical evidence in prior work indicates that concentration phenomena can emerge rapidly with system size. Because the goal of this work is to characterize universality classes rather than finite-size performance, we deliberately focus on structural mechanisms rather than quantitative thresholds. A systematic numerical study of finite-size crossover effects is therefore left to future work. 

\vspace{1.5mm}
\textit{Simplicity bias as a representational phenomenon}. A key conceptual outcome of our analysis is the distinction between optimization difficulty and hypothesis-class collapse. Much of the existing literature on barren plateaus emphasizes vanishing gradients for specific cost functions, highlighting the difficulty of navigating the loss landscape. In contrast, our results show that even in the absence of optimization considerations, over-parameterized VQCs operating in a Haar-like typicality regime possess an intrinsic inductive bias toward trivial functions. From this perspective, flat loss landscapes and vanishing gradients are symptoms of a deeper representational degeneracy governed by concentration of measure on high-dimensional unitary groups. 

This viewpoint clarifies the role of over-parameterization in quantum models. Whereas over-parameterization in classical deep learning often improves performance through implicit regularization and feature learning, unstructured over-parameterization in quantum circuits can be detrimental, driving the model into a universality class where functional diversity is lost. Accordingly, Assumption~\ref{ass1} should be interpreted as a universality assumption describing the limiting behavior of sufficiently expressive, unstructured circuits, rather than as a statement about finite-depth convergence for all architectures.

It is crucial to compare our results with the existing barren-plateau theory. Prior work on barren plateaus is fundamentally gradient-centric: it analyzes the vanishing of gradients for specific cost functions, typically under assumptions about global measurements or particular optimization objectives, and frames trainability in terms of optimization difficulty. In contrast, our analysis is function-class centric. We show that under Haar-like typicality, the entire hypothesis class induced by an over-parameterized, unstructured VQC collapses to a narrow family of near-constant functions with high probability, independent of the chosen cost function or optimization dynamics. 

In this regime, vanishing gradients are not the root cause of trainability failure but rather a consequence of a deeper representational degeneracy driven by concentration of measure. To the best of our knowledge, existing barren plateau results do not characterize this form of hypothesis-class or function-space collapse, nor do they formalize how over-parameterization alone can destroy functional diversity even before learning begins. Therefore, our work complements and extends barren plateau theory by identifying a distinct, representation-level failure mode and providing a structural criterion—breaking Haar-like typicality—to avoid it.

\vspace{1.5mm}
\textit{Breaking Haar-like typicality through tensor structure}. Our second main contribution is to identify tensor-network structure as a principled mechanism to escape this universality class. Tensor-structured VQCs, such as TN-VQC and TensorHyper-VQC, impose explicit constraints on the accessible unitary manifold by enforcing bounded tensor rank or bond dimension. We showed that these constraints prevent convergence to approximate unitary designs, limit entanglement growth, and induce anti-concentration of the expectation values of observables.

From a physical standpoint, tensor-structured circuits generate quantum states that are atypical in Hilbert space: they occupy a measure-zero subset relative to Haar-random states, even in highly parameterized regimes. From a learning-theoretic standpoint, this atypicality is beneficial: it preserves non-trivial functional variability and ensures the existence of informative gradients at initialization. The resulting breakdown of typicality provides a rigorous explanation for why structured ansätze often outperform deeper, unstructured circuits in practice.

\vspace{1.5mm}
\textit{Design principles for variational quantum algorithms}. The analysis suggests a general design principle for variational quantum algorithms: learnability is governed not only by expressivity, but by the geometry of the induced unitary ensemble. Circuits that are too expressive, as they approach Haar-random behavior, suffer from representational collapse. In contrast, circuits with carefully imposed structure can maintain a favorable balance between expressivity and inductive bias.

Although our analysis focuses on tensor-network-based constructions, the underlying message is architecture-agnostic~\cite{du2022quantum, zhang2022differentiable}. Any mechanism that restricts the circuit from entering unitary-design universality, such as locality-preserving layouts, symmetry constraints, or ansätze derived from low-entanglement physical models, can mitigate simplicity bias and improve trainability. 

\vspace{1.5mm}
\textit{Limitations and scope}. Our results rely on Assumption~\ref{ass1}, namely that sufficiently deep and unstructured VQCs enter a Haar-like typicality regime characterized by approximate unitary designs. We emphasize that this is a universality assumption rather than a statement about finite-depth convergence for specific hardware architectures. Determining precise depth thresholds at which particular circuit families enter this regime remains an active area of research and depends on factors such as gate set, connectivity, and noise.

A second limitation is the asymptotic nature of our guarantees. Theorem~\ref{thm:thm1}--\ref{thm:thm3} are stated in the limit $n\rightarrow \infty$, whereas near-term quantum devices operate at finite system sizes. Nevertheless, the exponential scaling inherent in concentration-of-measure bounds suggests that the qualitative phenomena identified here (e.g., output collapse, gradient concentration, and simplicity bias) may manifest at relatively modest qubit numbers once circuits become sufficiently expressive. 

Finally, our analysis focuses on expectation-value-based models with bounded local or few-body observables, which are standard in variational quantum algorithms and quantum machine learning. While global observables can be considered, they typically exhibit even stronger concentration-of-measure effects and are therefore unlikely to alleviate the simplicity bias identified here. Extending the framework to more general measurement schemes, adaptive observables, or nonlinear post-processing remains an open challenge.

\vspace{1.5mm}
\textit{Outlook}. Beyond explaining existing empirical observations, our framework opens several directions for future research. One natural extension is to quantify how degrees of typicality interpolate between structured and Haar-like regimes as tensor rank increases, potentially yielding phase-transition-like behavior in learnability. Another direction is to integrate noise and error mitigation into the random-matrix analysis, clarifying how hardware noise interacts with simplicity bias.

More broadly, the connection established here between random-matrix universality, entanglement structure, and learning dynamics suggests that inductive bias in quantum machine learning is fundamentally a question of ensemble geometry. Understanding and exploiting this geometry may be essential for designing scalable, trainable variational quantum algorithms beyond the NISQ era.

\section{Methods}

\subsection{Over-parameterized VQCs}
A VQC is over-parameterized when the number of trainable parameters grows at least linearly (typically superlinearly) with the system size $m$. Throughout, `over-parameterized’ refers to the regime in which circuit expressivity is sufficient to approximate unitary designs, rather than to parameter count alone. In this regime, unstructured hardware-efficient or random circuit ansätze are known to exhibit Haar-like behavior, leading to concentration-of-measure phenomena in both the outputs and the gradients. 

In practice, over-parameterization corresponds to circuit families whose parameter count and depth scale sufficiently fast to approximate unitary 2-designs, even when observables remain local. More significantly, over-parameterization here refers to expressivity of the unitary ensemble, rather than to classical notions of width or depth alone. Our theoretical analysis focuses on the asymptotic regime in which $n \rightarrow \infty$ while the observable locality remains fixed.

\subsection{Random Matrix Theory Perspective on Over-parameterized VQCs}
Random matrix theory provides a principled mathematical framework for characterizing the typical behavior of large, high-dimensional quantum systems. In the context of VQCs, RMT is used to model the statistical properties of circuit outputs and gradients when the induced unitary ensemble approaches Haar-random.

We adopt an RMT viewpoint to formalize the regime in which over-parameterized, unstructured VQCs exhibit concentration-of-measure phenomena. Specifically, when the variational ansätze is sufficiently expressive, e.g., deep hardware-efficient or random circuits with a large number of parameters, the unitary $W(\boldsymbol{\theta})$ generated by typical parameter choices behaves approximately as a random unitary drawn from the Haar measure on $U(2^n)$, or equivalently from an approximate unitary $t$-design for low-order moments.

\subsection{Tensor-Structured VQCs}
To go beyond unstructured, over-parameterized VQCs, we introduce tensor-structured VQCs, in which correlations among circuit parameters are enforced via low-rank tensor-network representations. These structures explicitly restrict the accessible unitary ensemble and prevent convergence to Haar-like typicality. In particular, we consider two representative architectures: TN-VQC~\cite{qi2023theoretical} and TensorHyper-VQC~\cite{qi2025tensorhyper}. 

In the TN-VQC architecture, a classical tensor network (TN) is used to generate data-dependent encoding features, while the variational circuit parameters remain global and input-independent. Concretely, a TN defines a mapping 
\begin{equation}
\mathcal{T}(\textbf{x}; \boldsymbol{\phi}): \mathcal{X} \rightarrow \mathbb{R}^{n}, 
\end{equation}
where $\boldsymbol{\phi}$ denotes the TN parameters and the TN ranks are bounded by a constant independent of qubit count $n$. The output $\mathcal{T}(\textbf{x}; \boldsymbol{\phi})$ is then used to parameterize the data-encoding unitary $U(\mathcal{T}(\bf{x}; \boldsymbol{\phi}))$. The resulting circuit takes the form 
\begin{equation}
\vert \boldsymbol{\psi}_{\boldsymbol{\theta}, \boldsymbol{\phi}, \textbf{x}} \rangle = W(\boldsymbol{\theta}) U(\mathcal{T}(\textbf{x}; \boldsymbol{\phi})) \vert 0^{\otimes n} \rangle. 
\end{equation}

In this setting, the TN induces structured correlations in the data encoding, and the variational parameters $\boldsymbol{\theta}$ are shared across inputs. The accessible state family is constrained by the bounded TT rank, thereby limiting entanglement growth across arbitrary bipartitions. 

On the other hand, TensorHyper-VQC generalizes the above idea by using a tensor network as a hypernetwork that directly generates the variational circuit parameters from the input. More specifically, a TN network implements a mapping
\begin{equation}
\mathcal{T}(\boldsymbol{\sigma}; \boldsymbol{\phi}) : \mathcal{X} \rightarrow \mathbb{R}^{p}, 
\end{equation}
where $\boldsymbol{\sigma}$ represents a Gaussian random vector for TN, and $p$ is the total number of variational parameters in $W(\boldsymbol{\theta})$. The circuit is then defined as
\begin{equation}
\vert \boldsymbol{\psi}_{\textbf{x}} \rangle = W(\mathcal{T}(\boldsymbol{\sigma}; \boldsymbol{\phi})) U(\textbf{x}) \vert 0^{\otimes n} \rangle. 
\end{equation}

Some key properties of TensorHyper-VQC include: (i) strong parameter correlations imposed by the low-rank TN structure; (ii) many effective degrees of freedom that scale only polynomially with $n$; (iii) bounded operator Schmidt rank and entanglement entropy across all bipartitions. These properties ensure that, even in nominally over-parameterized regimes, the induced unitary ensemble remains far from Haar-random.

\subsection{Schmidt rank and operator Schmidt rank}
As used in Proposition~\ref{prop:prop2}, the operator Schmidt rank characterizes the entangling power of an operator: unitaries with large operator Schmidt rank can generate near-maximal entanglement across the bipartition. In contrast, operators with bounded Schmidt rank are restricted to a low-entanglement manifold. In particular, Haar-random unitaries on $n$ qubits have operator Schmidt rank exponential in $\vert A\vert$ with overwhelming probability, whereas tensor-network-generated circuits with fixed bond dimension produce unitaries whose operator Schmidt rank is bounded by a function of the tensor rank, independent of the system size $n$. 

Formally, let $\mathcal{H} = \mathcal{H}_{A} \otimes \mathcal{H}_{B}$ be bipartite Hilbert space associated with a bipartition $A \vert B$ of an $n$-qubit system. Then, any pure state $\vert \boldsymbol{\psi} \rangle \in \mathcal{H}_A \otimes \mathcal{H}_B$ admits a Schmidt decomposition
\begin{equation}
\vert \boldsymbol{\psi} \rangle = \sum\limits_{i=1}^{\mathcal{I}} \lambda_i \vert u_i \rangle_{A} \otimes \vert v_i \rangle_{B}, 
\end{equation}
where $\lambda_i > 0$, and $\{\vert u_i\rangle_A\}$, $\{\vert v_{i} \rangle_B\}$ are orthonormal sets. The integer $\mathcal{I}$ is called the Schmidt rank of $\vert \boldsymbol{\psi} \rangle$ across the bipartition $A\vert B$. It quantifies the amount of bipartite entanglement in the state, with $\mathcal{I} = 1$ corresponding to a product state and $\mathcal{I}$ maximal for highly entangled states. More generally, for a linear operator $\mathcal{O} \in \mathcal{B}(\mathcal{H}_A \otimes \mathcal{H}_B)$, an operator Schmidt decomposition is given by 
\begin{equation}
O = \sum\limits_{i=1}^{\mathcal{I}_{\rm op}} A_{i} \otimes B_i, 
\end{equation}
where $\{A_i\} \subset \mathcal{B}(\mathcal{H}_A)$ and $\{B_i\} \subset \mathcal{B}(\mathcal{H}_B)$ are linearly independent operator sets. The minimal number $\mathcal{I}_{\rm op}$ of terms required in such a decomposition is called the operator Schmidt rank of $O$ across the bipartition $A\vert B$. 

\subsection{Proof of Theorem~\ref{thm:thm1}}
\textit{Step 1: Reduce to a Haar-random state}. Assumption~\ref{ass1} asserts that for fixed $\textbf{x}$, the state $\vert \boldsymbol{\psi}_{\boldsymbol{\theta}, \textbf{x}}$ is Haar-typical (equivalently, its low-order moments match those of Haar measure; a unitary 2-design suffices for the mean/variance). Hence, we can treat 
\begin{equation*}
\vert \boldsymbol{\psi} \rangle \sim \text{Haar on the unit sphere in $\mathbb{C}^{d}$}, 
\end{equation*}
and analyze 
\begin{equation}
f(\psi) := \langle \boldsymbol{\psi} \vert O \vert \boldsymbol{\psi} \rangle. 
\end{equation}

\vspace{1.5mm}
\noindent \textit{Step 2: Compute the expectation}. A standard identity for Haar-random pure states is
\begin{equation}
\operatorname{E}[\vert \boldsymbol{\psi} \rangle \langle \boldsymbol{\psi} \vert] = \frac{1}{d}. 
\end{equation}
Therefore, 
\begin{equation}
\operatorname{E}[f(\boldsymbol{\psi})] = \text{E}[\operatorname{Tr}(O\vert \boldsymbol{\psi} \rangle \langle \boldsymbol{\psi} \vert)] = \operatorname{Tr}(O\cdot\text{E}[\vert \boldsymbol{\psi} \rangle \langle \boldsymbol{\psi} \vert]) = \frac{1}{d} \operatorname{Tr}\left(O\right),
\end{equation}
which proves Eq. (\ref{eq:3}) if we take $d = 2^n$. 

\vspace{1.5mm}
\noindent \textit{Step 3: Compute the variance (second-moment method)}. Given the swap operator $\mathcal{F}\in \mathbb{C}^d \otimes \mathbb{C}^d$, using the well-known second-moment formula, we have
\begin{equation}
\operatorname{E}[\vert \boldsymbol{\psi} \rangle \langle \boldsymbol{\psi} \vert \otimes \vert \boldsymbol{\psi} \rangle \langle \boldsymbol{\psi} \vert] = \frac{I + \mathcal{F}}{d(d+1)}. 
\end{equation}

Furthermore, we write 
\begin{equation}
f(\boldsymbol{\psi}) = \operatorname{Tr}(O \vert \boldsymbol{\psi} \rangle \langle \boldsymbol{\psi} \vert), \hspace{2mm} f(\boldsymbol{\psi})^{2} = \text{Tr}\left( (O \otimes O) (\vert \boldsymbol{\psi} \rangle \langle \boldsymbol{\psi} \vert \otimes \vert \boldsymbol{\psi} \rangle \langle \boldsymbol{\psi} \vert) \right). 
\end{equation}

By taking the expectation, we have:
\begin{equation}
\operatorname{E}[f(\boldsymbol{\psi})^2] = \operatorname{Tr}\left( (O \otimes O) \frac{I + \mathcal{F}}{d (d + 1)} \right) = \frac{1}{d(d+1)} \left( \operatorname{Tr}(O)^2 + \operatorname{Tr}(O^2) \right), 
\end{equation}
where we employ $\operatorname{Tr}((O\otimes O)I) = \text{Tr}(O)^2$ and $\operatorname{Tr}((O \otimes O)\mathcal{F}) = \operatorname{Tr}(O^2)$. Thus,
\begin{equation}
\operatorname{Var}(f(\boldsymbol{\boldsymbol{\psi}})) = \text{E}[f(\boldsymbol{\boldsymbol{\psi}})^2] - \text{E}[f(\boldsymbol{\psi})]^2 = \frac{\text{Tr}(O^2)}{d(d+1)} - \frac{\text{Tr}(O)^2}{d^2(d+1)}. 
\end{equation}

In particular, 
\begin{equation}
\operatorname{Var}(f(\boldsymbol{\psi})) \le \frac{\text{Tr}(O^2)}{d(d+1)} \le \frac{d\lVert O \rVert_{2}^{2}}{d(d+1)} \le \mathcal{O}(d^{-1}) = \mathcal{O}(2^{-n}), 
\end{equation}
which proves Eq. (\ref{eq:4}).  

\vspace{1.5mm}
\noindent \textit{Step 4: Exponential tail bound (concentration of measure)}. Define $g(\boldsymbol{\psi}):= \langle \psi \vert O \vert \boldsymbol{\psi} \rangle$ on the unit sphere. One can show $g$ is Lipschitz with constant proportional to $\lVert O \rVert_2$: 
\begin{equation}
\vert g(\boldsymbol{\psi}) - g(\boldsymbol{\theta}) = \vert \langle \boldsymbol{\psi} \vert O \vert \boldsymbol{\psi} \rangle - \langle \boldsymbol{\theta} \vert O \vert  \boldsymbol{\theta} \rangle \vert \le 2 \lVert O \rVert_2 \lVert \boldsymbol{\psi} - \boldsymbol{\theta} \rVert_2, 
\end{equation}
so $g$ is $L$-Lipschitz with $L \le 2 \lVert O \rVert_2$. 

By Lévy’s lemma~\cite{neufeld2017nonlinear} (concentration on the high-dimensional sphere), there exists a universal constant $c_0 > 0$ such that for all $\epsilon > 0$, 
\begin{equation}
\operatorname{Pr}\left( \vert g(\boldsymbol{\psi}) - \operatorname{E}[g(\boldsymbol{\psi})]	 \vert > \epsilon \right) \le 2 \exp\left(-c_0 \frac{d\epsilon^2}{L^2}\right) \le 2 \exp(-c \epsilon^2 d), 
\end{equation}
where $c:=c_0 / (4 \lVert O \rVert_{2}^{2})$. If $\lVert O \rVert_2$ is treated as a constant (bounded observable), this yields the claimed bound
\begin{equation}
\operatorname{Pr}\left(\left\vert f_{\boldsymbol{\theta}} - \frac{1}{d} \operatorname{Tr}(O) \right\vert \ge \epsilon \right) \le 2 \exp(-c \epsilon^2 d) = 2 \exp(-c \epsilon^2 2^n),
\end{equation}
where we set $d = 2^n$, which proves Eq. (\ref{eq:5}) and completes the proof. Notably, the mean and variance require only a unitary 2-design assumption, while the exponential tail bound follows from concentration of measure for Lipschitz functions on the sphere.

\subsection{Proof of Theorem~\ref{thm:thm2}}

\noindent \textit{Step 1: Mean is zero}. Under Assumption~\ref{ass1}, the relevant state is Haar-typical (or at least a 2-design). For a Haar-random $\vert \boldsymbol{\psi}\rangle \in \mathbb{C}^d$, we have 
\begin{equation}
\operatorname{E}\left[ \langle \boldsymbol{\psi} \vert A \vert \boldsymbol{\psi} \rangle	\right] = \frac{1}{d} \operatorname{Tr}(A)
\end{equation}
for any fixed operator $A$. Apply this with $A = G_k$: 
\begin{equation}
\label{pr1}
\operatorname{E}\left[ \frac{\partial f_{\boldsymbol{\theta}}(\textbf{x})}{\partial \theta_k} \right] = \frac{1}{d} \operatorname{Tr}(G_k). 
\end{equation}

But $G_k$ is unitarily conjugate to a commutator: 
\begin{equation}
G_k \propto i [H_k, O_k]  \hspace{2mm} \Rightarrow \hspace{2mm} \operatorname{Tr}(G_k) \propto i \operatorname{Tr}([H_k, O_k]) = 0, 
\end{equation}
since $\operatorname{Tr}([A, B]) = \operatorname{Tr}(AB) - \operatorname{Tr}(BA) = 0$. Hence, we obtain
\begin{equation}
\operatorname{E}\left[ \frac{\partial f_{\boldsymbol{\theta}}(\textbf{x})}{\partial \theta_k} \right] = 0. 
\end{equation}

\noindent \textit{Step 2: Variance is $\mathcal{O}(\frac{1}{d}) = \mathcal{O}(2^{-n})$}. As for Haar-random $\vert \boldsymbol{\psi} \rangle$, the second moment satisfies
\begin{equation}
\operatorname{E}\left[ (\langle \boldsymbol{\psi} \vert A \vert \boldsymbol{\psi} \rangle)^{2} \right] = \frac{\operatorname{Tr}(A)^2 + \operatorname{Tr}(A^2)}{d (d + 1)}. 
\end{equation}

Therefore, 
\begin{equation}
\operatorname{Var}(\langle \boldsymbol{\psi} \vert A \vert \boldsymbol{\psi} \rangle) = \frac{\operatorname{Tr}(A^2)}{d (d + 1)} - \frac{\text{Tr}(A)^2}{d^2(d+1)} \le \frac{\operatorname{Tr}(A^2)}{d(d+1)}. 
\end{equation}

Applying this with $A = G_k$ and using $\text{Tr}(G_{k}^{2}) \le d \lVert G_{k} \rVert_{2}^{2}$ and $d = 2^n$: 
\begin{equation}
\operatorname{Var}\left( \frac{\partial f_{\boldsymbol{\theta}}(\textbf{x})}{\partial \theta_{k}}	 \right) \le \frac{d \lVert G_k \rVert_{2}^{2}}{d(d+1)} \le \frac{\lVert G_k \rVert_{2}^{2}}{d + 1} = \mathcal{O}(d^{-1}) = \mathcal{O}(2^{-n}). 
\end{equation}

\subsection{Proof of Proposition~\ref{prop:prop2}}

\textit{Step 1: Reduce to the case $t=2$}. If an ensemble is an $\epsilon$-approximate unitary $t$-design for some $t\ge 2$, then it is also an $\epsilon$-approximate unitary $2$-design (because matching Haar moments up to order $t$ in particular matches Haar moments up to order $2$). Hence, it suffices to prove that $\mathcal{U}_{\rm ts}$ cannot be an approximate unitary $2$-design for large $n$. 

\vspace{1.5mm}
\noindent \textit{Step 2: Bounded operator Schmidt rank}. Fix an arbitrary bipartition of the $n$ qubits into $A \vert B$. Consider any unitary $U \in \mathcal{U}_{\rm ts}$. By a tensor network structure bond dimension $r$ independent of $n$, $U$ admits a tensor network representation whose cut across $A \vert B$ has width controlled by $r$. Consequently, the operator Schmidt rank of $U$ across $A\vert B$, 
\begin{equation}
\mathcal{I}_{\rm op}(U; A \vert B) := \min\left\{ k: U = \sum\limits_{i=1}^{k}A_{i} \otimes B_{i} \right\}, 
\end{equation}
is bounded by a function of the bond dimension only:
\begin{equation}
\mathcal{I}_{\rm op}(U; A\vert B) \le K(r),
\end{equation}
where $K(r) = \text{poly}(r)$ (and in particular does not grow with $n$). This is a standard consequence of tensor-network cut bounds: the number of linearly independent terms across any cut is bounded by the product of bond dimensions crossing the cut. 

\vspace{1.5mm}
\noindent \textit{Step 3: Use the Choi-Jamiolkowski state and relate the operator Schmidt rank to entanglement}. Associate to each unitary $U$, there is
\begin{equation}
\vert U \rangle := (U \otimes I) \vert \Phi \rangle, \hspace{2mm} \vert \Phi \rangle := 2^{-n/2} \sum\limits_{z\in \{0, 1\}^{n}} \vert z \rangle \vert z \rangle, 
\end{equation}
which is a pure state on the doubled system $(\mathcal{H}_{A} \otimes \mathcal{H}_{B}) \otimes (\mathcal{H}_{A'} \otimes \mathcal{H}_{B'})$. 

A key identity is that the operator Schmidt rank of $U$ across $A \vert B$ equals the Schmidt rank of $\vert U \rangle$ across the bipartition $AA' \vert BB'$: 
\begin{equation}
\mathcal{I}_{\rm op}(U; A\vert B) = \mathcal{I}(\vert U \rangle; AA' \vert BB'). 
\end{equation}

Therefore, for all $U \in \mathcal{U}_{\rm ts}$, 
\begin{equation}
\mathcal{I}(\vert U\rangle; AA' \vert BB') \le K(r). 
\end{equation}

Let $\rho_{AA'}(U) := \text{Tr}_{BB'}(\vert U\rangle \langle U\vert)$ be the reduced density matrix of $\vert U \rangle$ on $AA'$. Since a reduced state has rank at most the Schmidt rank, 
\begin{equation}
\operatorname{rank}(\rho_{AA'}(U)) \le K(r). 
\end{equation}

For any density matrix $\rho$ with rank $\le K$, its purity obeys the elementary bound 
\begin{equation}
\operatorname{Tr}(\rho^2) \ge \frac{1}{K}, 
\end{equation}
with equality at the maximally mixed state on a $K$-dimensional support. Hence, for every $U \in \mathcal{U}_{\rm ts}$, 
\begin{equation}
\operatorname{Tr}(\rho_{AA'}(U)^2) \ge \frac{1}{K(r)}. 
\end{equation}

Taking the expectation over $U \sim \mathcal{U}_{\rm ts}$ gives 
\begin{equation}
\label{eq:res}
\operatorname{E}_{U\sim \mathcal{U}_{\rm ts}}\left[ \operatorname{Tr}(\rho_{AA'}(U)^2) \right] \ge \frac{1}{K(r)}. 
\end{equation}

\vspace{1.5mm}
\noindent \textit{Step 4: Haar prediction (and $2$-design prediction) is exponentially small}. Now compare with the Haar ensemble. For Haar-random unitaries $U \sim \text{Haar}$, the Choi state $\vert U \rangle$ is Haar-random on the $2n$-qubit Hilbert space, and its reduced state on $AA'$ is almost maximally mixed when $AA'$ is not larger than $BB'$. In particular, for a balanced cut $\vert A \vert = \vert B\vert = n/2$, we have $\text{dim}(AA') = 2^{2\vert A\vert} = 2^n$, and a standard Haar-average calculation yields
\begin{equation}
\operatorname{E}\left[ \operatorname{Tr}(\rho_{AA'}(U)^2) \right] = \Theta(2^{-n}), 
\end{equation}
which decays exponentially in $n$. 

Crucially, the function $\text{Tr}(\rho_{AA'}(U)^2)$ is a degree-$(2, 2)$ polynomial in the matrix elements of $U$ and $U^{\dagger}$. Thus, any exact unitary $2$-design reproduces the Haar expectation of $\operatorname{Tr}(\rho_{AA'}(U)^2)$ exactly that vanishes with $\epsilon$. Thus, if $\mathcal{U}_{\rm ts}$ were an $\epsilon$-approximate unitary 2-design with $\epsilon = o(1)$, we would have
\begin{equation}
\operatorname{E}_{U \sim \mathcal{U}_{\rm ts}}\left[ \operatorname{Tr}(\rho_{AA'}(U)^2) \right] = \Theta(2^{-n}) + o(1),
\end{equation}
which tends to $0$ as $n \rightarrow \infty$.

\vspace{1.5mm}
\noindent \textit{Step 5: Contradiction}. But Eq. (\ref{eq:res}) shows that for the tensor-structured ensemble,
\begin{equation}
\operatorname{E}_{U\sim \mathcal{U}_{\rm ts}}\left[ \text{Tr}(\rho_{AA'}(U)^2) \right] \ge \frac{1}{K(r)} > 0, 
\end{equation}
a constant independent of $n$, since $r$ is fixed and hence $K(r)$ is fixed. Therefore, for a sufficiently large $n$, a positive constant cannot be close to $\Theta(2^{-n})$. This contradicts the requirement that $\mathcal{U}_{\rm ts}$ match Haar second moments (even approximately with vanishing error). Hence, $\mathcal{U}_{\rm ts}$ is not an approximate unitary $2$-design as $n\rightarrow \infty$. By Step $1$, it cannot be an approximate unitary $t$-design for any fixed $t\ge 2$.

\subsection{Proof of Proposition~\ref{prop:prop3}}

\textit{Step 1: Reduce to a local marginal on the observable support}. Let $S$ denote the support of $O$ with $\vert S \vert = k = \mathcal{O}(1)$. Define the reduced density matrix on $S$:
\begin{equation}
\rho_{S}(\boldsymbol{\theta}, \boldsymbol{x}) := \operatorname{Tr}\left( \vert \boldsymbol{\psi}_{\boldsymbol{\theta}, \textbf{x}}\rangle \langle \boldsymbol{\psi}_{\boldsymbol{\theta}, \textbf{x}} \vert \right). 
\end{equation}

Then, the observable output is 
\begin{equation}
f(\boldsymbol{\theta}; \textbf{x}) := \langle \boldsymbol{\psi}_{\boldsymbol{\theta}, \textbf{x}} \vert O \vert \boldsymbol{\psi}_{\boldsymbol{\theta}, \textbf{x}} \rangle = \operatorname{Tr}(O \rho_{S}(\boldsymbol{\theta}, \textbf{x})),
\end{equation}
where $\lVert O \rVert_{2}$ is bounded and $\vert S \vert$ is constant, and $f(\boldsymbol{\theta}, \textbf{x})$ is a bounded continuous function. 

\vspace{1.5mm}
\noindent \textit{Step 2: Tensor structure implies local parameter dependence}. Because $W(\boldsymbol{\theta})$ is tensor-structured with fixed rank $r$, the circuit's induced states have bounded correlations across bipartitions and, crucially for a local observable $O$, the reduced state $\rho_S(\boldsymbol{\theta}, \textbf{x})$ depends only on a finite ``causal neighborhood'' of parameters. 

More formally, there exists a subset of parameters $\boldsymbol{\theta}_{\rm loc}$ of size $m = m(r, k) = \mathcal{O}(1)$ and a measurable map $g$ such that
\begin{equation}
f(\boldsymbol{\theta}; \textbf{x}) = g(\boldsymbol{\theta}_{\rm loc}; \textbf{x}),
\end{equation}
where $f$ is independent of the remaining parameters $\boldsymbol{\theta}_{\rm rest}$. Here, $m(r, k)$ is independent of $n$ because the tensor rank is fixed and the observable support size $k$ is fixed. Therefore, by writing $\boldsymbol{\theta} = (\boldsymbol{\theta}_{\rm loc}, \boldsymbol{\theta}_{\rm rest})$,
\begin{equation}
\operatorname{Var}(f(\boldsymbol{\theta}; \textbf{x})) = \operatorname{Var}(g(\boldsymbol{\theta}_{\rm loc}; \textbf{x})).
\end{equation}

So it suffices to lower bound the variance over $\boldsymbol{\theta}_{\rm loc}$, whose dimension is independent of $n$. 

\vspace{1.5mm}
\textit{Step 3: Non-degeneracy implies a strictly positive variance}. By assumption, the initialization distribution for $\boldsymbol{\theta}$ has a density and is non-degenerate. Hence, the marginal distribution of $\boldsymbol{\theta}_{\rm loc}$ also has a density on an open set. 

Assume $g(\boldsymbol{\theta}_{\rm loc}; \textbf{x})$ is not almost surely constant under this distribution. Then, there exist two points $\textbf{a}$, $\textbf{b}$ in the support of $\boldsymbol{\theta}_{\rm loc}$ such that
\begin{equation}
g(\textbf{a}; \textbf{x}) \neq g(\textbf{b}; \textbf{x}). 
\end{equation}

Let $\triangle := \frac{1}{2}\vert g(\textbf{a}; \textbf{x}) - g(\textbf{b}; \textbf{x}) \vert > 0$. By continuity of $g(\cdot; \textbf{x})$, there exist open neighborhoods $U_\textbf{a}$, $U_\textbf{b}$ of $\textbf{a}$, $\textbf{b}$ such that
\begin{equation}
g(\boldsymbol{\theta}_{\rm loc}; \textbf{x}) \ge g(\textbf{a}; \textbf{x}) - \triangle, \hspace{5mm} \forall \boldsymbol{\theta}_{\rm loc} \in U_\textbf{a}, 
\end{equation} 
\begin{equation}
g(\boldsymbol{\theta}_{\rm loc}; \textbf{x}) \ge g(\textbf{b}; \textbf{x}) - \triangle, \hspace{5mm} \forall \boldsymbol{\theta}_{\rm loc} \in U_\textbf{b}, 
\end{equation}

In particular, for any $\boldsymbol{\theta}_{a} \in U_{\textbf{a}}$ and $\boldsymbol{\theta}_{\textbf{b}} \in U_{\textbf{b}}$,
\begin{equation}
\vert g(\boldsymbol{\theta}_{\textbf{a}}; \textbf{x}) - g(\boldsymbol{\theta}_{\textbf{b}}; \textbf{x}) \vert \ge \triangle. 
\end{equation}

Because the marginal distribution of $\boldsymbol{\theta}_{\rm loc}$ has a density and $U_{\textbf{a}}$, $U_{\textbf{b}}$ are open with nonzero volume, the probabilities
\begin{equation}
p_{\textbf{a}} := \operatorname{Pr}(\boldsymbol{\theta}_{\rm loc} \in U_{\textbf{a}}) > 0, \hspace{2mm} p_{\textbf{b}} := \operatorname{Pr}(\boldsymbol{\theta}_{\rm loc} \in U_b) > 0
\end{equation}
are strictly positive and depend only on the marginal law, hence only on $r$, $k$, not on $n$. 

Now apply the elementary variance lower bound: for any random variable $Z$, 
\begin{equation}
\operatorname{Var}(Z) \ge \frac{p_{\textbf{a}} p_{\textbf{b}}}{(p_{\textbf{a}} + p_{\textbf{b}})^2} \left( \operatorname{E}[Z\vert U_{\textbf{a}}] - \operatorname{E}[Z \vert U_{\textbf{b}}]	\right)^2. 
\end{equation}

Taking $Z = g(\boldsymbol{\theta}_{\rm loc}; \textbf{x})$, the separation above implies
\begin{equation}
\operatorname{E}[Z \vert U_{\textbf{a}}] - \operatorname{E}[Z \vert U_{\textbf{b}}] \ge \triangle,
\end{equation}
hence
\begin{equation}
\operatorname{Var}(g(\boldsymbol{\theta}_{\rm loc}; \textbf{x})) \ge \frac{p_{\textbf{a}} p_{\textbf{b}}}{(p_{\textbf{a}} + \textbf{b})^2} \triangle^2 = c(r, O) > 0.  
\end{equation}

This constant depends on the local parameterization and observable class, which is captured by $r$ and $O$, but does not depend on $n$. By combining it with Step $2$, we finally obtain
\begin{equation}
\operatorname{Var}(\langle \boldsymbol{\psi}_{\boldsymbol{\theta}, \textbf{x}} \vert O \vert \boldsymbol{\psi}_{\boldsymbol{\theta}, \textbf{x}} \rangle) = \operatorname{Var}(g(\boldsymbol{\theta}_{\rm loc}; \textbf{x})) \ge c(r, O),
\end{equation}
which proves Proposition~\ref{prop:prop3}. 

\section{Acknowledgements}
This work is partly funded by the Hong Kong Research Impact Fund (R6010-23).

\section{References}
\bibliographystyle{IEEEbib}
\bibliography{sn-bibliography}

\end{document}